\documentclass[aps,prl,twocolumn,superscriptaddress]{revtex4}

\usepackage{amsmath}
\usepackage{amssymb}
\usepackage{bm}
\usepackage{epsfig}
\usepackage{graphicx}
\newcommand{\be}{\begin{equation}}
\newcommand{\ee}{\end{equation}}
\begin{document}

\title{Quantum Adiabatic Computation With a Constant Gap is Not Useful in
One Dimension}
\author{M.~B.~Hastings}
\affiliation{Center for Nonlinear Studies and Theoretical Division, Los Alamos
National Laboratory, Los Alamos, NM 87545}
\begin{abstract}
We show that it is possible to use a classical computer
to efficiently simulate the adiabatic evolution
of a quantum system in one dimension with a constant spectral gap, starting
the adiabatic evolution from a known initial product
state.  The proof relies
on a recently proven area law for such systems, implying the existence
of a good matrix product representation of the ground state, combined with
an appropriate algorithm to update the matrix product state as the
Hamiltonian is changed.  This implies that adiabatic evolution
with such Hamiltonians is not
useful for universal quantum computation.  Therefore,
adiabatic algorithms
which are useful for universal quantum computation
either require a spectral gap tending to zero or need to be implemented
in more than one dimension (we leave open the question of the computational
power of adiabatic simulation with a constant gap in more than one dimension).
\end{abstract}
\maketitle

There are many different models for quantum computation.  The most
standard approach is the gate model, combined with appropriate error
correction to deal with decoherence\cite{gate}.  Other approaches
include measurement based quantum computer\cite{mbqc}, topological
quantum computing\cite{tqc}, and adiabatic
quantum computation\cite{aqc}.  Adiabatic quantum computation
is very natural
because one can imagine slowly changing the Hamiltonian following a path
in parameter space, starting from some simple Hamiltonian with a known
ground state, and arriving at some final Hamiltonian whose ground state
encodes the solution of a difficult optimization problem\cite{aqcopt}.

While adiabatic quantum computation has been shown to allow for universal
quantum computation\cite{aduniv},
and hence is equivalent in
its computational power, the fault tolerant approach and related
threshold theorems\cite{thresh} have not been generalized to
adiabatic quantum computation.  
Instead, one can rely on the spectral gap in the Hamiltonian to protect
against errors.  The spectral gap is also interesting in adiabatic
quantum computation because the time required to perform the computational
scales with the inverse spectral gap.  It has at least been shown\cite{jfs}
that one can produce a constant gap against {\it local} noise on 1 and
2 qubits, but has never been shown that one can produce a constant
gap against all excitations.

Therefore, it is of great interest to determine if universal adiabatic
quantum computation can be performed in systems with a spectral
gap of order {\it unity}.  In a sense, topological
quantum computation provides a means of performing universal adiabatic
quantum computation with constant gap,
by adiabatically changing the Hamiltonian to drag defects around each other.
However,
this topological approach relies on having a large ground state degeneracy.
In this paper we show that, at least in one dimension, adiabatic quantum
computation in systems with a unique ground state and a constant
spectral gap is not useful for quantum computation as it can be
simulated efficiently on a classical computer.

{\it Main Result---}
We consider the following problem.
Consider a parameter-dependent
Hamiltonian $H(s)=\sum_i h_{i,i+1}(s)$, with $h_{i,i+1}(s)$
having support on sites $i,i+1$ and with $\Vert h_{i,i+1}(s) \Vert \leq J$,
so that interactions are nearest neighbor.
Let there be $N$ sites, and assume that each site has a Hilbert space
dimension $D$ which is ${\cal O}(1)$.
Assume that for all $s$ with
$0\leq s \leq s_{max}$ we have a spectral gap $\Delta E$ with
$J/\Delta E$ being ${\cal O}(1)$.   
Finally, assume that $\Vert \partial_s h_{i,i+1} \Vert \leq J$; this last
requirement simply sets some scale for how large $s$ is.  In general,
if we have $\Vert \partial_s h_{i,i+1} \Vert\leq X$, for any constant
$X>J$, we can rescale $s\rightarrow sX/J$ and $s_{max}\rightarrow s_{max}X/J$,
and with this rescaled $s$ the requirement
$\Vert \partial_s h_{i,i+1} \Vert \leq J$ becomes satisfied.
Assume that $H(0)$ has a known product ground state.  Consider
any observable $O$ which is a product of operators supported on a single
site.

Our main result is that it is
possible to compute the expectation value of $O$ in the
ground state of $H(s_{max})$ to any desired accuracy
$\epsilon_O$
by an algorithm that takes a computational time $T$ given by
\be
T=\exp[{\cal O}(D^{{\cal O}(\omega)})]
{\cal O}(N (J/\Delta E) s_{max} (N/\epsilon_0)^{\omega}),
\ee
on a {\it classical} computer,
where the exponent $\omega$ equals
\be
\omega=
{\cal O}\Bigl(\ln (D) J/\Delta E\Bigr).
\ee
For any operator $P$ supported
on a constant number of sites, $n$, we can write $P$ as a sum over $D^{2n}$
different product operators $O$.  Therefore, the ability to approximate
product operators implies the ability to approximate
general operators on a constant number of sites.
This result implies that physical quantities such as the ground
state energy of $H(s_{max})$ can be approximated to within accuracy
$1/{\rm poly}(N)$
in
polynomial time on a classical computer.  

We rely heavily
on the area law for one-dimensional systems with gapped Hamiltonians\cite{1darea} proven recently.
Let $\psi^0(s)$ denote the ground state of $H(s)$.
Given that
$H(s)$ satisfies the conditions above including
$J/\Delta E$ being ${\cal O}(1)$, the area law implies
that, for any $\epsilon>0$, we can approximate
$\psi^0(s)$ by
a matrix product state $\psi_{mps}(s)$ 
such that
\be
\label{areares}
|\psi^{mps}(s)-\psi^0_s|^2 \leq \epsilon,
\ee
and such that $\psi_{mps}(s)$ has bond dimension
\be
\label{ispoly}
k={\cal O}(N/\epsilon)^{\omega} \exp[{\cal O}(D^{{\cal O}(\omega)})].
\ee
The bond dimension is polynomial in $N/\epsilon$, but may be
doubly exponentially large in $J/\Delta E$; this general upper bound
applies to all one-dimensional systems while specific cases in practice
have not required such a large bond dimension.
While the area law implies the existence of a good matrix product approximation
to the ground state, it does not imply that we can efficiently find the correct
matrix product state.  To solve this problem, we construct a sequence of
matrix product approximations to the ground states along the entire path
in parameter space, as we now explain (the explanation of how to do this
given in \cite{1darea} was not correct, and we now give a full
description of the correct construction).
Given such a matrix product state, we can efficiently calculate
expectation values of product operators such as $O$.

{\it The Algorithm---}
At $s=0$, the
ground state is assumed to be a known product state, and hence is
a matrix product state of bond dimension $k=1$.  We
break the adiabatic evolution for $s=0$ to $s=s_{max}$ into
a sequence of polynomially many discrete steps, such that $s$ increases
by a small amount in each step.  
Specifically, let $a_{max}$ be the smallest integer larger than
$4 N J s_{max}/\Delta E$, and let $\delta=s_{max}/a_{max}$ so
$\delta\leq \Delta E/4 N J$.
We break the adiabatic evolution into $a_{max}$ different
discrete steps of size $\delta$, and
we set $s_a=a \delta$, for $a=0,1,2,...,a_{max}$.
We have $a_{max}={\cal O}(N(J/\Delta E)s_{max})$.
We fix a maximum bond dimension $k_{max}$, by setting $k_{max}=k$ for
$k$ given by Eq.~(\ref{ispoly}) with $\epsilon=1/{\rm poly}(N)$; the correct
choice of $\epsilon$ to obtain a given error is given after Eq.~(\ref{givene1}).
Let $\psi^0_a$ denote the
ground state of $H(s_a)$.
For $a=0$, we represent the ground state $\psi^0_a$ exactly as a matrix product
state $\psi^{mps}_a$.
Our algorithm proceeds iteratively, taking a series of $a_{max}$ steps,
such that on the $a$-th step it computes a matrix product
state
$\psi^{mps}_a$ of bond dimension at most $k_{max}$ which
is a good approximation to
$\psi^0_a$.
We do this using the following algorithm:
\begin{itemize}
\item[{\bf 1:}] Initialize $\psi^{mps}_0$ to the known initial product state.

\item[{\bf 2:}] {\bf For} $a=1$ to $a_{max}$ {\bf do}
\begin{itemize}
\item[{\bf 2a:}] Compute a matrix product
state $\psi^t_a$ from the state $\psi^{mps}_{a-1}$
as described in the section {\it Improving the Approximation}.  
$\psi^t_a$
will be a good approximation to $\psi^0_a$, but its bond dimension $k'$
may be
larger than $k_{max}$ by a factor polynomial in $N$ and $\epsilon$.

\item[{\bf 2b:}] Compute the state $\psi^{mps}_a$ from the state $\psi^t_a$
by a truncation procedure described in {\it Truncation Error Bounds}.  The
state $\psi^{mps}_a$ will be a matrix product state of bond dimension
at most $k_{max}$.
\end{itemize}
\end{itemize}

We define the error after each step, $\epsilon_a$, by
\begin{eqnarray}
\label{ov2}
|\langle \psi^{mps}_a,\psi^0_a\rangle|^2 = 1-\epsilon_a.
\end{eqnarray}
The correctness of the algorithm is based on the inductive assumption
that after $a-1$ steps, we have
\be
\label{indassumpt}
\epsilon_{a-1} \leq {\rm min}(\Delta E/12 N J,1/99),
\ee
so that $\psi^{mps}_{a-1}$ is a good approximation to $\psi^0_{a-1}$.
As we show in Eq.~(\ref{firsterr}),
since
the difference
$|s_{a}-s_{a-1}|$ is less than or equal to $\Delta E/4 N J$,
the state
$\psi^{mps}_{a-1}$ is also a good approximation to
$\psi^{0}_{a}$.  We then use the result (\ref{firsterr}) and the
spectral gap and locality of the Hamiltonian to
construct a state $\psi^t_{a}$ which is a better
approximation to $\psi^0_{a}$
at the cost of increasing the bond dimension above
$k_{max}$ as described two sections later.
We then truncate $\psi^t_{a}$ to a state $\psi^{mps}_{a}$
with bond dimension $k_{max}$, and we use the area law to
bound errors in this truncation.

Both the computational effort required
to compute $\psi^t_a$
and the bond dimension $k'$
will be proportional to
$(N/\epsilon_a)^{{\cal O}(\omega)}$.
Since it is possible, given a matrix product state of
bond dimension $k$, to compute expectation values of observables which are
products of single site operators in a time polynomial in $k$, this
shows the main result.

{\it Difference Between $\psi^{mps}_{a-1}$ and $\psi^0_a$---}
Assume by induction that after $a-1$ steps we have a good approximation $\psi^{mps}_{a-1}$ to
$\psi^0_{a-1}$, namely that Eq.~(\ref{indassumpt}) holds.
Since $|s_{a}-s_{a-1}|\leq
\Delta E/4 N J$, we have
\be
\label{diffHbnd}
\Vert H(s_{a})-H(s_{a-1})\Vert \leq \Delta E/4.
\ee

From Eq.~(\ref{diffHbnd}), 
\be
\Bigl|
\langle \psi^0_{a-1}, H(s_a) \psi^0_{a-1}\rangle-
\langle \psi^0_{a-1}, H(s_{a-1}) \psi^0_{a-1}\rangle
\Bigr|\leq \Delta E/4.
\ee
Let $E^0_a$ denote the ground state energy of $H(s_a)$.  We have
$|E^0_{a}-E^0_{a-1}|\leq \Delta E/4$.  Therefore,
\be
\langle \psi^0_{a-1}, H(s_a) \psi^0_{a-1}\rangle-E^0_a
\leq \Delta E/2.
\ee
Since $H(s_{a})$ has a spectral gap $\Delta E$,
\be
\label{ov1}
|\langle \psi^0_a,\psi^0_{a-1}\rangle|^2\geq 1/2.
\ee

From Eq.~(\ref{ov2}), the angle $\theta_1$ between vectors
$\psi_{a-1}^{mps}$ and $\psi_{a-1}^0$ obeys $\cos(\theta_1)^2\geq 1-\epsilon_{a-1}$.
From Eq.~(\ref{ov1}), the angle $\theta_2$ between vectors
$\psi_{a-1}^0$ and $\psi_a^0$ obeys $\cos(\theta_2)^2\geq 1/2$.
Therefore, we can bound the angle $\theta$ between vectors $\psi_{a-1}^{mps}$
and $\psi_a^0$ by $\theta\leq \theta_1+\theta_2$ with
$\cos(\theta_1+\theta_2)^2=[\cos(\theta_1)\cos(\theta_2)-\sin(\theta_1)\sin(\theta_2)]^2
\geq \cos(\theta_1)^2\cos(\theta_2)^2-2\cos(\theta_1)\sin(\theta_1) \cos(\theta_2)\sin(\theta_2)\geq (1-\epsilon_{a-1})(1/2)-\sqrt{\epsilon_{a-1}}\geq 1/4$.
where the last inequality follows because
$\epsilon_{a}\leq 1/99$
by (\ref{indassumpt}).
Therefore, since 
$|\langle \psi^{mps}_{a-1},\psi^0_{a}\rangle|^2= \cos(\theta)^2$,
\begin{eqnarray}
\label{firsterr}
|\langle \psi^{mps}_{a-1},\psi^0_{a}\rangle|^2
\geq  1/4.
\end{eqnarray}

{\it Improving the Approximation---}
We now explain step {\bf 2a}, 
which uses a state
$\psi^{mps}_{a-1}$ which satisfies (\ref{firsterr})
to construct, for any desired $\epsilon$, a state $\psi^t_{a}$ such that $|\langle \psi^t_a,\psi^0_a\rangle|^2\geq 1-\epsilon$,
with a bond dimensional and computational cost of order
$(N/\epsilon)^{{\cal O}(\omega)}$.
In practice, we would prefer simpler constructions of the state $\psi^t_a$, as
explained in the discussion.

Let the states $\psi^k_a$ be
a complete basis of eigenstates of $H(s_a)$ with energies $E^k_a$.
From Eqs.~(\ref{diffHbnd},\ref{indassumpt}) and the fact that
$\Vert H(s_a)\Vert \leq NJ$, after $a-1$ steps it is possible for
the algorithm to compute
$E^0_a$ to an accuracy of $\Delta E/4+(\Delta E/12 NJ)(NJ)=
\Delta E/3$\cite{rna}.  In the following, assume that the
algorithm estimates $E^0_a$ to be zero (if this is not the case,
shift the energies $E^k_a$ by the estimate
of $E^0_a$).

Consider the state
\begin{eqnarray}
\label{slof}
\phi_a&=&
\frac{\Delta E}{\sqrt{2\pi q}}
 \int {\rm dt} \exp[-(\Delta E t)^2/2q] \exp[i H(s_a) t]
\psi^{mps}_{a-1} 
\nonumber
\\
&=& \sum_{k} |\psi^k_{a}\rangle \exp[-q(E^k_a/\Delta E)^2/2] \; \langle \psi^k_a|\psi^{mps}_{a-1}\rangle,
\end{eqnarray}
where $q$ is a number we choose below to be of order $\log(1/\epsilon)$,
where $\epsilon$ is the error estimate below in (\ref{diff0}).
We will first show that the difference between
$\phi_a/|\phi_a|$ and $\psi^0_a$ is exponentially small in $q$
and then we will show how to approximate $\phi_a$ by a matrix product
state, giving the desired matrix product approximation to $\psi^0_a$.
Let $\psi^{mps}_{a-1}=A_a \psi^0_{a}+B_a \psi^\perp_a$
where $\langle \psi^0_a,\psi^\perp_a \rangle=0$. Note
that by Eq.~(\ref{firsterr}), we have
$|A_a|^2\geq 1/4$.  Let
$\phi_a=A'_a \psi^0_{a}+B'_a \phi^\perp_a$ with
$\langle \psi^0_a,\phi^\perp_a\rangle=0$.
The idea of the integration over time is to
approximately project onto the ground state;
the projection of $\phi_a$ on any state $\psi^k_a$
is reduced by a factor of
$\exp[-q(E^k_a/\Delta E)^2/2]$ compared
to the projection of $\psi^{mps}_{a-1}$ onto $\psi^k_a$.
Therefore, since
$|E^0_a|\leq \Delta E/3$,
we have $A_a'\geq (1/4) \exp(-q/18)$.  However, since all other
states have an energy at least $\Delta E$ above the ground
state, and hence an energy at least $2\Delta E/3$ above
zero, we have $B_a'\leq \exp(-2q/9)$.  Thus, we can guarantee that
the normalized state $\phi_a/|\phi_a|$ is within
error $\epsilon'$ of $\psi^0_a$ for any $\epsilon'$ by
choosing $q$ logarithmically large in $N$ and $\epsilon'$.

We now show how to approximate $\phi_a$ by a matrix product state.
First, replace the integral over all $t$ between $-\infty$ and $+\infty$
by
a finite integral, from
$t=-t_{max}$ to $t=+t_{max}$ with $t_{max}=99 q/\Delta E$.
The error in making
this replacement is of order $\exp[-(\Delta E t_{max})^2/2q]=\exp(-99^2 q/2)$.  Also, replace
the continuous integral over $t$ by a discrete sum over different
times $t_i$.  The bound
on the operator norm of $H(s_a)$, $\Vert H(s_a) \Vert \leq N J$,
implies that $|\partial_t \exp[i H(s_a) t]\psi_{a-1}^{mps}|\leq NJ$,
and so
we can approximate the integral by a sum with
an error $\epsilon'$ using $n_{sum}={\cal O}((t_{max} \Delta E/\sqrt{2\pi q})(N J t_{max}/\epsilon'))={\cal O}(\sqrt{q} N J t_{max}/\epsilon')$ terms in the sum.

We now approximate the sum of these $n_{sum}$ states by
a matrix product state with polynomial bond dimension.
It was shown\cite{osb} using Lieb-Robinson bounds\cite{lr1,lr2}
that the state $\exp[i H(s_a) t_i] \psi^{mps}_{a-1}$
can be approximated to error $\epsilon'$ by a matrix product state with a bond
dimension $\exp({\cal O}(t_i J \ln D))k_{max}$ times
a function of $N/\epsilon'$ which grows slower than
any power.  Since $t_i$ is of order $(1/\Delta E)\ln(N/\epsilon')$,
the state $\exp[i H(s_a) t_i] \psi^{mps}_{a-1}$ can be approximated
by a state with
bond dimension of order $(N/\epsilon')^{{\cal O}(\omega)}k_{max}$\cite{coinc}.  The sum over
$n_{sum}$ different matrix product states is still a matrix product
state, with a bond dimension which is $n_{sum}$ times as large.
Let the normalized sum of matrix product states be $\psi^t_a$.
Therefore for any $\epsilon$ we can choose $q$ such that
\be
\label{diff0}
|\psi^t_{a}-\psi^0_{a}|^2\leq \epsilon,
\ee
and such that $\psi^t_a$
is a matrix
product state with bond dimension polynomial in $N$ and $1/\epsilon$
and such that computing $\psi^t_a$ requires only polynomial
computational effort.

{\it Truncation Error Bounds---}
We now bound the truncation error introduced in {\bf 2b}.
By the area law, the ground state $\psi^0_{a}$ obeys
\be
\label{diff}
|\chi_a-\psi^0_{a}|^2\leq \epsilon,
\ee
for some $\chi_a$ which is a matrix product state of bond dimension $k$,
and thus from (\ref{diff0},\ref{diff}) we have
\be
\label{diff2}
|\chi_a-\psi^t_{a}|^2\leq 4\epsilon.
\ee
Pick any bond and do a Schmidt decomposition of $\psi^t_{a}$ across that bond,
writing
$\psi^t_{a}=\sum_{\alpha} A(\alpha) \psi_L(\alpha) \otimes \psi_R(\alpha)$.
Order the Schmidt coefficients so that $|A(\alpha)|$ is decreasing as
$\alpha$ increases.
Then, from Eq.~(\ref{diff2}),
\be
\label{bds}
\sum_{\alpha>k} |A(\alpha)|^2\leq
4\epsilon.
\ee

We now define $\psi_{a}^{mps}$.  As shown in \cite{cv}, the bound
on Schmidt coefficients in Eq.~(\ref{bds}) implies that there
exists a matrix product state $\psi_a^{mps}$ with bond dimension
$k$ such that
\be
\label{mpserr}
|\psi^{mps}_a-\psi^t_a|^2\leq 8(N-1)\epsilon.
\ee
The construction in \cite{cv} is defined in terms of the
matrices which are the parameters of the matrix product state;
an equivalent construction is to
define, for each $i=2,...,N$, the operator $P_{i,N}$ to project onto the
$k$ largest Schmidt coefficients of the reduced density matrix
$\psi^t_{a}$ on sites $i,...,N$.
Then, let
$\psi^{mps}_a$ be defined by
\be
\psi^{mps}_a=Z^{-1} P_{N-1,N} P_{N-2,N} ... P_{2,N} \psi^t_{a},
\ee

By Eqs.~(\ref{diff},\ref{mpserr}),
\be
|\psi^{mps}_a-\psi^0_a|^2\leq
\epsilon+
8(N-1)\epsilon+2\sqrt{8(N-1)}\epsilon.
\ee
Therefore,
\be
|\langle \psi^{mps}_{a},\psi^0_a\rangle|^2\geq 1
-\epsilon_a,
\ee
with
\be
\label{givene1}
\epsilon_{a+1}=
\epsilon+
8(N-1)\epsilon+2\sqrt{8(N-1)}\epsilon.
\ee
By picking $\epsilon$ small enough, we can ensure that (\ref{indassumpt}) is
true after $a$ steps, given that it was true after $a-1$ steps, and
we can make $\epsilon_{a_{max}}$ polynomially small at a polynomial computational cost.
This completes the error estimates.

{\it Discussion---}
This work raises several natural questions in the field of Hamiltonian
complexity.  First, what happens in more than one dimension?  We do
not know of a general proof of an area law in more than one dimension, and it is quite conceivable
that adiabatic simulation in two dimensions with a constant gap and a unique ground
state {\it is} computationally
universal.

Second, the algorithm we have chosen is not very practical.  While the construction of
the state $\psi^t_a$ requires only polynomial time, and hence this work suffices to show
in principle that adiabatic simulation with a constant gap is not useful in one dimension, in
practice we prefer to construct matrix product states with as low a bond dimension as possible.
This is why low order Trotter-Suzuki approaches are popular, such as the
TEBD algorithm\cite{tebd,tebd2}.  Therefore, it is of great interest to prove that the following
natural algorithm always suffices to approximate the ground state:
make the spacing
$s_{a}-s_{a-1}$ very small (polynomially small in $N$) so that the state $\psi^{mps}_{a-1}$ is
polynomially close to approximating the ground state $\psi^{0}_a$.  Then,
in step {\bf 2a}, set
$\psi^t_a=(1-H(s_a)/\Vert H(s_a)\Vert) \psi^{mps}_{a-1}$, or perhaps instead 
$\psi^t_a=\exp(-H_{even}(s_a) \tau)\exp(-H_{odd}(s_a)\tau)\psi^{mps}_{a-1}$ where $\tau$ is some small
quantity and $H_{even,odd}(s_a)$ represent the terms of $H(s_a)$ on the even and odd bonds
respectively.  Finally, do the truncation step {\bf 2b} as before.
Such a proof would require a much more accurate analysis of the error in the
truncation step.

Finally, it is important to understand the role that the adiabatic evolution plays in
our result.  The area law implies the existence of a good matrix product state approximation
to the ground state.  This implies that the following decision problem is in NP: given a
Hamiltonian $H$ with interaction strength $J$ and local Hilbert space dimension both
${\cal O}(1)$ and with nearest neighbor interactions, and given the
promises that $H$ has a unique ground state with inverse spectral gap which is ${\cal O}(1)$,
and that the ground state energy $E_0$ is either $\leq 0$ or $\geq 1/{\rm poly}(N)$, decide
whether the ground state energy is indeed less than zero.  Such a problem is in NP
because the matrix product state guaranteed by the area law acts as a witness, in computer
science language, or as a variational state, in physics language.  However, we have no
guarantee that we can efficiently find such a state.  Indeed, it has been shown that
there are problems with a ground state which has a polynomial bond dimension as a matrix
product state for which computing this matrix product state is NP-hard\cite{nphard}.
This result \cite{nphard} holds for systems with an inverse polynomial spectral gap, rather than those
with a constant gap as we consider here, so it is not certain what the difficulty is of
finding matrix product states for the systems we consider, but it may indeed by a difficult
problem.  Thus, the ability to {\it follow} the state along the adiabatic change in the
Hamiltonian is useful precisely because it allows one to be sure of locating a good
matrix product approximation to the ground state.

{\it Acknowledgments---} I thank D. Aharonov and G. Freedman
for many useful discussions, and R. Hanson for comments on a draft.
This work was
supported by U. S. DOE Contract No. DE-AC52-06NA25396.


\begin{thebibliography}{99}
\bibitem{rna} We ignore error due to the finite numerical
precision of
real number arithmetic on a computer.
Such error can be taken into
account straightforwardly if desired.
\bibitem{coinc} The same exponent, $\omega$,
up to to constant factors, appears in \cite{1darea} and here,
because both constructions
start with state with non-zero overlap with the
ground state and bootstrap this to construct a state with a better
overlap.
\end{thebibliography}
\end{document}